\documentclass[reprint,amsmath,amssymb,aps,prb,floatfix,showkeys]{revtex4-2}

\usepackage{graphicx}
\usepackage{dcolumn}
\usepackage{bm}
\usepackage{hyperref}

\begin{document}

\title{Optical orientaion of excitons in hybrid metal-semiconductor nanostructures}

\author{A. V. Korotchenkov}
    \email{coalex@mail.ioffe.ru}
\author{N. S. Averkiev}

\affiliation{Ioffe Institute, Politechnicheskaya 26, St. Petersburg, 194021, Russian Federation\\}

\date{\today}

\begin{abstract}
We demonstrate the possibility of the optical orientation of excitons in the near field of the metal grating that covers a semiconductor nanostructure.
Excitons generated this way have the wave vector greater than the wave vector of the incident radiation.
We suggest that optical orientation method is applicable to study the fine structure and kinetics of the hot excitons in semiconductor quantum well.
\end{abstract}

\maketitle

\section{Introduction}

Principles of optical orientation and alignment, which were initially
developed for atomic gases, are widely employed in the spectroscopy of
excitons in semiconductors \cite{Pikus1982}, \cite{Planel1984}.
As well as in bulk crystals \cite{Gamarts1977}, these methods allow investigating the fine structure of excitons in semiconductor nanostructures and estimating the exciton lifetime and spin relaxation time \cite{Dzhioev1997}, \cite{Smirnova2023}.
Normally one generates the excitons with polarized light at resonance frequency
\(\omega_{0}\), and measures the polarization of luminescence due to direct optical transitions.
This way, either free excitons with small wave vector \(\bm{K} \approx 0\), which they obtain from the incident light, or the localized excitons are studied.
In bulk semiconductors, orientation of hot excitons that have significant
\(\bm{K}\) and kinetic energy is only attainable through indirect transitions accompanied by the optical phonon emission \cite{Bir1976}.
On the other hand, it is possible to fabricate a metallic grating that supports
the surface plasmon polaritons (SPPs) on top of a semiconductor
heterostructure and preserve the exciton resonance \cite{Vasa2008}.
In the resulting hybrid nanostructure, excitons acquire the wave vector and
polarization (spin) of SPPs that propagate along the grating \cite{Akimov2021}.
This means, in principle, that one could study the properties of excitons with non-zero wave vectors by exciting them through the grating and measuring their polarized luminescence.

In this paper, we consider orientation of excitons in a hybrid nanostructure that consists of a semiconductor quantum well (QW) and a square grating of metal nanoparticles.
First, we find the electric field scattered by the grating, approximating the nanoparticles by discrete electric dipoles.
Polarization of the near field they produce inside the QW is quite different from polarization of the incident electromagnetic wave.
Next, we calculate the exciton generation matrix and solve the kinetic equation to obtain the stationary exciton density matrix.
At this stage, we consider the effect of the stationary homogeneous magnetic field applied perpendicularly to the QW plane.
Since the grating allows simultaneously generating excitons that propagate in different directions, we take into account both exciton spin and exciton
momentum relaxation.
Finally, we calculate the polarization of secondary radiation of the excitons facilitated by the grating.
We aim to find out how the luminescence polarization varies with the applied magnetic field, and demonstrate that hybrid nanostructures make it possible to study the kinetic characteristics of the hot excitons.

\section{Generation of excitons through the
grating}
\begin{figure}[b]
	\includegraphics[width=.4\textwidth]{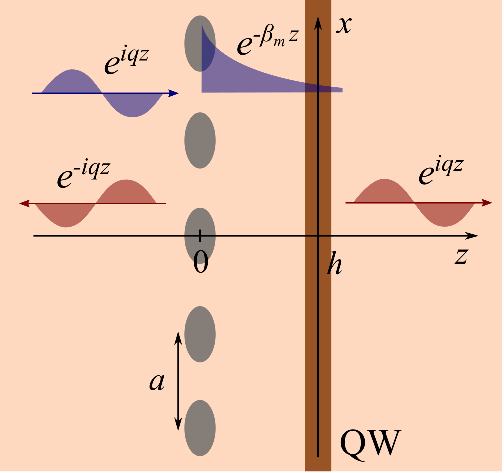}
	\caption{Schematic of nanostructure consisting of a semiconductor quantum well (QW) and a short-period grating of metal nanoparticles (NPs).}
    \label{Fig:1}
\end{figure}
\begin{figure}
    \includegraphics[width=0.4\textwidth]{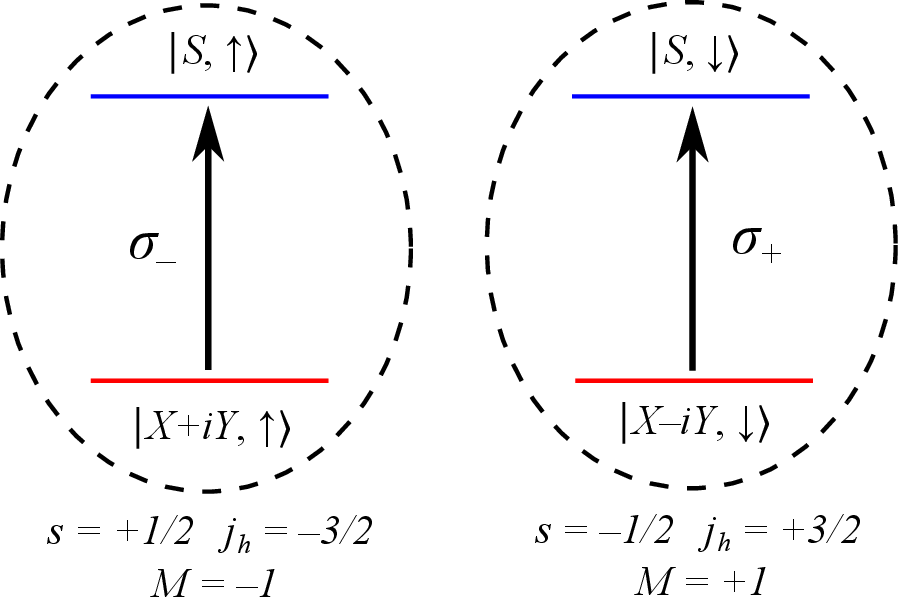}
    \caption{Optically active exciton states formed by conduction band electrons and heavy holes in the valence band.
    Electron states are labeled with the Bloch function $S$ or $X \pm iY$ and spin $\uparrow$ or $\downarrow$.
    Excitons with spin $M=-1$ or $+1$ interact with the left or right circularly polarized light correspondingly.}
    \label{Fig:2}
\end{figure}

Consider a structure shown in Fig.~\ref{Fig:1}, which contains a semiconductor quantum well (QW) near the grating of metallic nanoparticles.
We assume the photon energy of the incident radiation close to the energy of the excitonic transition in QW, and the distance \(h\) between the grating and the QW to not exceed the wavelength in semiconductor.
Then the excitons are generated by the near field of metallic grating, which results from scattering of the incident beam represented by the plane wave with certain polarization tensor.
We suppose that the grating does not affect the intrinsic properties of the excitons in QW.
To capture the essence of optical orientation and alignment effects, we employ a simple model of exciton transitions shown in Fig.~\ref{Fig:2}.
Due to the size quantization of charge carriers in QW, excitons in the ground state are formed by the heavy holes with angular momentum projection \(j_{h} = \pm 3/2\) on the growth axis \(z\), and by electrons with spin projection \(s = \pm 1/2\).
States with projection of total spin on the growth axis \(\pm 1\) are optically active, while states with spin
projection \(\pm 2\) are optically inactive and therefore can be neglected.
Optically active (bright) and inactive (dark) excitons can hybridize due to exchange interaction, which plays significant role in orientation of localized excitons \cite{Dzhioev1998}, but we do not take this effect into account here.
According to the selection rules shown in Fig.~\ref{Fig:2}, light with circular polarization \(\sigma_{+}\) generates excitons with spin projection\(\ M = + 1\), while \(\sigma_{-}\)-polarized light generates excitons with \(M = - 1\).
Then the secondary radiation of excitons partially retains the circular polarization of incident beam.
Linearly polarized light (we determine polarization in the QW plane \(xy\)) generates the superposition of exciton states \(\pm 1\) such that the total spin is oriented along the \(x\) or \(y\) axis, which is called the optical alignment effect.
The radiation from recombination of excitons in this state is also linearly polarized.
An external magnetic field oriented along \(z\) axis splits the states with \(M = + 1\) and reduces the coherence of superposition states, which decreases the linear polarization of exciton radiation.
Measurements of the dependence of luminescence polarization on the strength or direction of magnetic field provide estimations of exciton relaxation times and of \emph{g}-factor value for excitons.

This simple picture is suitable for a structure with no grating, where incident light generates only excitons with near zero wave vectors \(\bm{K}\) in the QW plane.
Now we consider a square lattice of conducting nanoparticles, in which the surface plasmon frequency lies in the same range as the frequency of exciton transition in the QW.
The lattice is chosen square so that it does not introduce additional polarization to the incident radiation and does not alter the polarization of the light emitted by the excitons with \(\bm{K} = 0\). 
To calculate the field of nanoparticles, we use the discrete dipoles approximation \cite{Chen2017}, which yields the total electric field in the following form
\begin{gather}    
    \bm{E}\left( \bm{\rho},z \right) = \bm{E}^{0}\exp(iqz) + \sum_{\bm{m}}^{}\bm{E}_{\bm{m}}\exp\left( i\bm{b_m}\cdot\bm{\rho} + iq_{z,\bm{m}}|z| \right),\label{Eq:1}\\
   \bm{E_m}= \frac{iq^{2}b^{2}}{2\pi n^{2}q_{z,\bm{m}}} \hat{G}\left( \bm{b_m} + {\rm sgn} z\ q_{z,\bm{m}} \bm{e}_{z} \right) \mathcal{X}(\omega)\;\bm{E}^{0}. \nonumber
\end{gather}
This equation contains the sum of the scattered waves that have the wave vector projections \(\bm{q}_{\parallel}\) on the structure plane \(\bm{\rho} = (x,y)\) equal to the reciprocal lattice vectors \(\bm{b_m} = b\left( m_{x}\bm{e}_{x} + m_{y}\bm{e}_{y} \right)\), where \(b = 2\pi/a\), \(a\) is the lattice period.
In case of the short-period lattice that we intend to consider, the length of the
vectors \(\bm{b}_{\bm{m \neq 0}}\) exceeds the wave number \(q = n\omega/c\) of light propagating in semiconductor with refractive index \(n\).
Therefore, the wave numbers \(q_{z,\bm{m}} = (q^{2} - \bm{b_m}^{2})^{1/2} = i\beta_{\bm{m}}\) are purely imaginary, and the corresponding waves decay exponentially
along the direction \(z\) perpendicular to the lattice.
The amplitudes of scattered waves in the dipole approximation are determined by the
effective polarizability of nanoparticles in the lattice \(\mathcal{X}(\omega)\) and the tensor \(\hat{G}\left( \bm{q} \right) = \bm{e}_s\otimes\bm{e}_s + \bm{e}_p\otimes\bm{e}_p\) that describes the radiation of an electric dipole.
This notation involves the tensor product of polarization vectors \(\bm{e}_s = \lbrack \bm{e}_z \times \bm{q}_{\parallel} \rbrack/q_{\parallel}\) and \(\bm{e}_p = \lbrack \bm{e}_s \times \bm{q} \rbrack/q\), whose definitions are extended to the region of evanescent waves \(q_{\parallel} > q(\omega)\).

Since the wave vector component parallel to the QW is conserved in optical transitions, the field of the lattice (\ref{Eq:1}) can produce excitons with wave vectors \(\bm{K} = \bm{b_m}\).
At the same time, the exciton transition energy \(\mathcal{E}_{\rm exc}(K) \approx \mathcal{E}_{\rm exc}(0) + \hbar^{2}K^{2}/2M_{\rm exc}\), where \(M_{\rm exc}\) is the exciton effective mass, must be close to the energy of incoming photons. 
Optical transitions are characterized by the matrix element of the interaction operator
\(V_{\bm{K},M} = \langle {\rm exc}_{\bm{K},M}| - \int\hat{\bm{d}}(\bm{r}) \cdot \bm{E}\ d^{3}\bm{r}|0\rangle\),
which we calculate using the smooth envelope approximation for the exciton wave function, see section 2.7.5 of the book \cite{Ivchenko2005}.
The density of the transition dipole moment
$\langle {\rm exc}_{\bm{K},M} | \hat{\bm{d}}\left( \bm{r} \right) | 0 \rangle = ie/(m_{0}\omega_{0})\ \bm{p}_{M}F_{\bm{K}}^\ast(\bm{r},\bm{r})$
is expressed via the wave function envelope with coinciding electron and hole coordinates, the interband matrix element of the momentum operator \(\bm{p}_{M}\), charge \(e\) and mass \(m_{0}\) of the free electron.
For simplicity, we assume that the exciton spin states are independent of its wave vector \(\bm{K}\), therefore they are labeled with the same projection \(M\) as the states with \(\bm{K} = 0\).
Given that the envelope function can be written as
\(F_{\bm{K}}\left( \bm{r},\bm{r} \right) = S^{- 1/2}\exp\left( i\bm{K} \cdot \bm{\rho} \right)\Phi(z - h)\),
where \(S\) is the QW area, after integrating the dipole moment density with the electric field (\ref{Eq:1}) we obtain
\begin{widetext}
    \begin{equation}
    \label{Eq:2}
    V_{\bm{K},M} = (2\pi)^{2}S^{- \frac{1}{2}}\sum_{\bm{m}}^{}{\delta\left( \bm{K} - \bm{b_m} \right)}\int{\exp\left( iq_{z,\bm{m}}z \right)}\Phi^\ast(z - h)dz \;
    \frac{ie}{m_{0}\omega_{0}}\bm{p}_{M}\left\lbrack \delta_{\bm{m},0} + \frac{iq^{2}b^{2}}{2\pi n^{2}q_{z,\bm{m}}}\mathcal{X}{\hat{G}}_{\bm{b_m}} \right\rbrack\bm{E}^{0}.
\end{equation}
\end{widetext}
The delta function \(\delta(\bm{K}-\bm{b_m})\) indicates the conservation of the wave vector component in the QW plane, which assumes the values of the reciprocal lattice vectors \(\bm{b_m}\).
The aforementioned selection rules involving the exciton spin correspond to the interband matrix element \(\bm{p}_{M} = p_{\rm cv}\ \bm{e}_M^\ast\), where
\(\bm{e}_{\pm} = (\bm{e}_x\pm i\bm{e}_y)/\sqrt{2}\) are unit vectors of circular polarization.
The overlap integral with the function \(\Phi(z)\) entering equation (\ref{Eq:2}) was estimated in \cite{Averkiev2019} for thin QWs, and can be deemed virtually independent of the wave number \(q_{z,\bm{m}}\).

Now we can obtain the generation matrix, that is, the rate of increase in the exciton density matrix due to the absorption of light scattered on the grating.
We restrict to the diagonal components in terms of the exciton wave vector~\(\bm{K}\)
\begin{equation}
    \label{Eq:3}
    g_{MM'}\left( \bm{K} \right) = 2\pi\hbar^{-1} V_{\bm{K},M}V_{\bm{K}\bm{,}M'}^\ast\;\delta(\hbar\omega - \mathcal{E}_{\rm exc}(\bm{K})).
\end{equation}
Photons of energy \(\hbar\omega = \mathcal{E}_{\rm exc}(0)\) generate excitons with \(\bm{K} = 0\), and the related matrix elements \(V_{0, \pm} \sim \bm{e}_{\pm}^\ast \cdot \bm{E}^{0} = E_{\pm}^{0}\) are proportional to the circularly polarized components of the incident wave.
In the presence of a square lattice of symmetric nanoparticles, the amplitude of the incident wave \(\bm{E}^{0}\) is multiplied by the transmittance coefficient
$t(\bm{K}=0) = 1 + iq^{2}b^{2}\mathcal{X}/(2\pi n^{2}q_{z,\bm{m}})$,
which does not depend on the light polarization.
Therefore, the lattice does not affect the optical orientation (and polarized luminescence) of excitons with \(\bm{K} = 0\), and one can use the phenomenological theory from \cite{Bir1973} to outline this effect.
To calculate the generation of excitons by partially polarized light, the products of the amplitudes \(E_{\alpha}^{0}{E_{\beta}^{0}}^\ast\) in equation (\ref{Eq:3}) are replaced with the components \(d_{\alpha\beta}^{0}\) of the polarization matrix of incident radiation, which is convenient to represent in the basis of circular polarization vectors \(\bm{e}_{\pm}\) in the form
\(d_{MM'}^{0} = I_0/2\ (\delta_{MM'}\  + \bm{\sigma}_{MM'} \cdot \bm{\mathcal{P}}^{0})\).
It involves the scalar product of the vector of Pauli matrices
\(\bm{\sigma}=(\sigma_{x}\bm{,}\sigma_{y}\bm{,}\sigma_{z})\)
and the vector of Stokes parameters of incident light
\(\bm{\mathcal{P}}^{0} = (\mathcal{P}_{l}^{0},\mathcal{P}_{l'}^{0},\mathcal{P}_{c}^{0})\).
These are the degree of linear polarization in \(x,y\)-axis, the degree of linear polarization in rotated by \(45^{\circ}\) \(x',y'\)-axis, and the degree of the circular polarization.
Then the exciton generation matrix at \(\bm{K} = 0\) coincides, up to a factor, with the polarization matrix in the circular basis~\cite{Bir1973}
\begin{eqnarray}
    g_{MM'}(\bm{K} ) &=& \frac{g_{0}I_{0}}{2}
    \begin{pmatrix}
        1 + \mathcal{P}_{c}^{0} & \mathcal{P}_{l}^{0} - i\mathcal{P}_{l'}^{0} \\
        \mathcal{P}_{l}^{0} + i\mathcal{P}_{l'}^{0} & 1 - \mathcal{P}_{c}^{0}
    \end{pmatrix}\nonumber\\
    &&\times \delta(\bm{K})\ \delta\left( \hbar\omega - \mathcal{E}_{\rm exc}(0) \right).
    \label{Eq:4}
\end{eqnarray}
It follows from equation (\ref{Eq:4}) that the average spin of excitons in the moment of generation \(\bm{M}^{0} = {\rm Tr}(g\bm{\sigma})/{\rm Tr}(g)\) is equal to the vector \(\bm{\mathcal{P}}^{0}\) that determines the polarization of incident light.

Combining equations (\ref{Eq:2}) and (\ref{Eq:3}), we obtain the generation matrix for excitons with \(\bm{K}\neq 0\)
\begin{widetext}
\begin{gather}
    \label{Eq:5}
    g_{MM'}( \bm{K}\bm{\neq}0) = \sum_{\bm{m}}{g_{\bm{m}} \left(\bm{e}_{M}^\ast \hat{G}_{\bm{b_m}}\bm{E}^{0}\right) \left( \bm{e}_{M'}\hat{G}_{\bm{b_m}}^\ast \bm{E}^{0\ast} \right)\ \delta\left( \bm{K}-\bm{b_m} \right)\ \delta\left( \hbar\omega - \mathcal{E}_{\rm exc}( \bm{b_m}) \right)},\\
    g_{\bm{m}} = (2\pi)^{3}\hbar^{- 1}\left( \frac{e p_{\rm cv}} {m_0\omega_0} \right)^{2} \left| \int\exp(iq_{z,\bm{m}}z) \Phi^\ast(z - h)dz \right|^2\left(\delta_{\bm{m},0} + \frac{q^2 b^2}{2\pi n^2 q_{z,\bm{m}}} \mathcal{X} \right)^{2}.\nonumber
\end{gather}
\end{widetext}
The contributions with \(\bm{m} \neq 0\) are nonzero solely due to the lattice, and the spin polarization of generated excitons is determined by the scattered waves \(\bm{E_m}\sim\hat{G}(\bm{b_m} )\bm{E}^{0}\).
Consider the incoming photons with the energy \(\hbar\omega = \mathcal{E}_{\rm exc}(0) + \hbar^{2}b^{2}/2M_{\rm exc}\), such that excitons having wave vectors
\(\bm{K} = \pm b\bm{e}_x\) or \(\pm b\bm{e}_y\) are generated due to the scattering (in the first diffraction order, see Fig. 3).
For these values of \(\bm{K}\) the nonzero components of tensor \(\hat{G}(\bm{K})\) are \(G_{xx}(\pm b\bm{e}_x ) = G_{yy}(\pm b\bm{e}_y) = \eta\)
and \(G_{yy}(\pm b\bm{e}_x) = G_{xx}(\pm b\bm{e}_{y}) = 1\),
where the parameter \(\eta = 1 - b^{2}/q^{2}\) is responsible for the change of polarization of scattered field compared to the polarization of light incident on the grating. Namely, the component of the electric field \(\bm{E}^{0}\) parallel to the wave vector \(\bm{K}\) is scaled \(\eta\) times relative to the perpendicular component of the field.
We consider this circumstance in (\ref{Eq:5}) and replace the products of field components with the polarization matrix to find
\begin{widetext}
\begin{gather}
    g_{MM'}(\bm{K}) = \sum_{\bm{b_m}}{\tilde{g}_{MM'}(\bm{b_m})}\;\delta\left( \bm{K} - \bm{b_m} \right)\; I_{0}\delta\left(\hbar\omega - \mathcal{E}_{\rm exc}(b) \right), \nonumber\\
    {\tilde{g}}_{MM'}\left( \pm b\bm{e}_{x} \right) = \frac{g_{1}}{4}
    \begin{pmatrix}
        1 + \eta^{2} + \mathcal{P}_{l}^{0}\left( \eta^{2} - 1 \right) + 2\eta\mathcal{P}_{c}^{0} & \eta^{2} - 1 + \mathcal{P}_{l}^{0}\left( \eta^{2} + 1 \right) - 2i\eta\mathcal{P}_{l'}^{0} \\
        \eta^{2} - 1 + \mathcal{P}_{l}^{0}\left( \eta^{2} + 1 \right) + 2i\eta\mathcal{P}_{l'}^{0} & 1 + \eta^{2} + \mathcal{P}_{l}^{0}\left( \eta^{2} - 1 \right) - 2\eta\mathcal{P}_{c}^{0}
    \end{pmatrix}, \nonumber\\
    {\tilde{g}}_{MM'}\left( \pm b\bm{e}_{y} \right) = \frac{g_{1}}{4}
    \begin{pmatrix}
        1 + \eta^{2} + \mathcal{P}_{l}^{0}\left( 1 - \eta^{2} \right) + 2\eta\mathcal{P}_{c}^{0} & 1 - \eta^{2} + \mathcal{P}_{l}^{0}\left( \eta^{2} + 1 \right) - 2i\eta\mathcal{P}_{l'}^{0} \\
        1 - \eta^{2} + \mathcal{P}_{l}^{0}\left( \eta^{2} + 1 \right) + 2i\eta\mathcal{P}_{l'}^{0} & 1 + \eta^{2} + \mathcal{P}_{l}^{0}\left( 1 - \eta^{2} \right) - 2\eta\mathcal{P}_{c}^{0}
    \end{pmatrix}. \label{Eq:6}
\end{gather}
\end{widetext}
Equation (\ref{Eq:6}) implies that excitons having four allowed wave vector values are produced at different rates \({\rm Tr}g(\bm{K})\) and with different average spins.
Note that the product \(I_{0}\delta( \hbar\omega - \mathcal{E}_{\rm exc}(K))\) in equations (\ref{Eq:4}) - (\ref{Eq:6}) originates from the consideration of monochromatic light, and in general case it should be replaced by the spectral intensity of radiation \(I_{0}(\mathcal{E}_{\rm exc}(K))\) with required photon energy.

\section{Exciton density matrix}
To calculate the characteristics of the recombination radiation of excitons, we require the stationary value of their density matrix \(\rho_{MM'}(\bm{K})\) determined by the equation
\begin{equation}
    \label{Eq:7}
    - \frac{i}{\hbar}\left\lbrack \mathcal{H}_{B},\rho \right\rbrack - \frac{\rho}{\tau} - \frac{\rho - \left\langle \rho \right\rangle_{\varphi}}{\tau_{p}} + \left( \frac{\partial\left\langle \rho \right\rangle_{\varphi}}{\partial t} \right)_{s.r.} + g\left( \bm{K} \right) = 0.
\end{equation}
We assume here that the density matrix has spin indices \(M = \pm 1\), but it is diagonal in the momentum \(\hbar\bm{K}\) of exciton motion in the QW plane.
This approximation is valid when the elastic scattering of excitons by impurities is considered, and the density matrix is averaged over random locations of scattering centers \cite{Kohn1957}.
The first term in equation (\ref{Eq:7}) is responsible for the evolution of the exciton spin in an external magnetic field.
We consider the case of longitudinal magnetic field \(B_{z}\), in which the Hamiltonian \(\mathcal{H}_{B} = \frac{1}{2}\hbar\Omega\ \sigma_{z}\) leads to the splitting \(\hbar\Omega = g_{\parallel}\mu_{\rm B}B_{z}\) of states with \(M = \pm 1\).
The second term in (\ref{Eq:7}) describes the recombination of excitons characterized by the lifetime~\(\tau\).
The third term represents the relaxation of exciton momentum due to elastic scattering, which brings the density matrix towards the value \(\langle \rho \rangle = \frac{1}{2\pi}\int_{0}^{2\pi}{\rho(\bm{K}) d\varphi}\) averaged over the direction of exciton momentum \(\hbar\bm{K}\).
We assume the momentum relaxation time \(\tau_{p}\) to be the shortest, which substantially simplifies the general theory \cite{Gamarts1977}.
The term \(\left(\partial\langle\rho\rangle/\partial t \right)_{\rm s.r.}\) describing the spin relaxation process is taken into account in the simplest form
\begin{gather}
    \frac{\partial}{\partial t}\left( \langle\rho_{+ +}\rangle - \langle \rho_{- -}\rangle \right)_{\rm s.r.} = - \frac{\langle\rho_{+ +} \rangle - \langle\rho_{- -}\rangle}{\tau_{s1}},\nonumber\\
    \frac{\partial}{\partial t}\left\langle \rho_{+ -} \right\rangle_{\rm s.r.} = - \frac{\langle \rho_{+ -}\rangle}{\tau_{s2}},\quad
    \frac{\partial}{\partial t}\left\langle \rho_{- +} \right\rangle_{\rm s.r.} = - \frac{\langle \rho_{- +}\rangle}{\tau_{s2}}. \label{Eq:8}
\end{gather}
Thus the averaged \(z\)-component of the exciton spin \(\langle M_{z}\rangle = \left(\langle\rho_{+ +}\rangle - \langle\rho_{- -}\rangle\right)/{\rm Tr}\langle\rho\rangle\ \) is considered to vanish in time \(\tau_{s1}\), while perpendicular components of exciton spin \(\langle M_{x}\rangle = \left(\langle \rho_{+ -}\rangle + \langle\rho_{- +}\rangle\right)/{\rm Tr}\langle\rho\rangle\) and \(\langle M_{y}\rangle = i\left(\langle\rho_{+ -}\ \rangle - \langle\rho_{- +}\ \rangle\right)/{\rm Tr}\langle\rho\rangle\) vanish in time \(\tau_{s2}\).
Such phenomenological picture is not restricted to specific spin relaxation mechanism, as it refers to the spin averaged over the exciton momentum orientation.

In this section, we do not take into account the exciton energy relaxation due to inelastic scattering, and seek the spin density matrix for a constant values of energy and, naturally, \(|\bm{K}|\).
The secondary radiation of such excitons has the same frequency as the incident light.
\begin{figure}[b]
    \centering
    \includegraphics[width=0.4\textwidth]{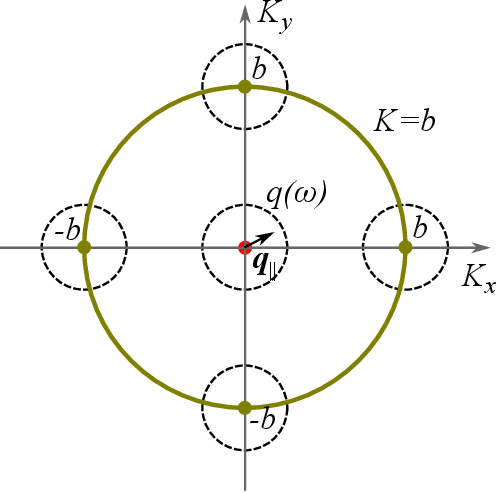}
    \caption{Distribution in \(\bm{K}\)-space of the excitons
    generated by the light with the frequency \(\omega = \omega_{\rm exc}(b)\). Dashed circles indicate the regions \(\left| \bm{K} - \bm{b_m} \right| = |\bm{q}_{\parallel}| < q(\omega)\) of ``bright'' excitons that can emit photons with in-plane wave vector \(\bm{q}_{\parallel}\).}
    \label{Fig:3}
\end{figure}

To solve equation (\ref{Eq:7}), we average it over the wave vector \(\bm{K}\) orientation, obtain the average density matrix and substitute it back into equation (\ref{Eq:7}).
It is convenient to represent the spin density matrix in the form
\(\rho_{MM'} = N/2 \left( \delta_{MM'} + \bm{\sigma}_{MM'} \cdot \bm{M} \right)\),
where \(N = {\rm Tr}\rho\) is the number of excitons with a given wave vector, and \(\bm{M}={\rm Tr}(\rho\bm{\sigma})/N\) is the average angular momentum per exciton.
A similar expansion was previously introduced for the matrix of light polarization and the matrix of exciton generation at \(\bm{K} = 0\).
The angular momentum components \(M_{x,y}\) and \(M_{z}\) evolve independently, even as the spin relaxation (\ref{Eq:8}) is included.
Only the term \(- i\hbar^{- 1}\lbrack\mathcal{H}_{B},\rho \rbrack = \Omega N ( \sigma_{y}M_{x} - \sigma_{x}M_{y})\) mixes the components in the QW plane, which corresponds to the precession of spin in magnetic field \(B_{z}\).
Then after averaging equation (\ref{Eq:7}) we find
\begin{gather}
    \langle N \rangle = \tau\ {\rm Tr}\langle g \rangle,\quad
    \langle N M_{z} \rangle = \tau_{1}\left\langle g_{+ +} - g_{- -} \right\rangle,\nonumber\\
    \left\langle N(M_{x} - iM_{y}) \right\rangle = \frac{\tau_{2}}{1 + i\Omega\tau_{2}}\ 2\left\langle g_{+ -} \right\rangle,\label{Eq:9}\\
    \left\langle N(M_{x} + iM_{y}) \right\rangle = \frac{\tau_{2}}{1 - i\Omega\tau_{2}}\ 2\left\langle g_{- +} \right\rangle.\nonumber    
\end{gather}
Here \(\tau_{i}^{- 1} = \tau^{- 1} + \tau_{si}^{- 1}\) are the summary decay rates of the total exciton spin components.
Next, we substitute the obtained averaged spin components back into (\ref{Eq:7}) and designate \(\tau_\ast^{-1} = \tau^{- 1} + \tau_{p}^{- 1}\) to find the result
\begin{gather}
    N = \tau_\ast\left( {\rm Tr}g + \frac{\tau}{\tau_{p}}{\rm Tr}\langle g \rangle \right),\nonumber\\
    \label{Eq:10}
    NM_{z} = \tau_{\ast}\left( g_{++} - g_{--} + \frac{\tau_{1}}{\tau_{p}}\left\langle g_{++} - g_{--} \right\rangle \right),\\
    N\left(M_{x} - iM_{y}\right) = \frac{2\tau_{\ast}}{1 + i\Omega\tau_{\ast}} \left( g_{+-} + \frac{\tau_{2}}{\tau_{p}}\frac{\left\langle g_{+-} \right\rangle}{1 + i\Omega\tau_{2}} \right)\nonumber\\
    = N\left( M_{x} + iM_{y} \right)^{\ast}.\nonumber
\end{gather}
Since the exciton generation matrix (\ref{Eq:6}) has the form
\(g(\bm{K}) = \sum_{|\bm{b_m}|=b}{I_{0}\;\tilde{g}(\bm{b_m})\delta(\bm{K}-\bm{b_m})}\),
its averaged over directions value is \(\left\langle g(\bm{K}) \right\rangle = \left(\tilde{g}(b\bm{e}_{x}) + \tilde{g}(b\bm{e}_{y})\right)\ I_{0}/(\pi b)\ \delta(K - b)\).
Therefore, the resulting density (\ref{Eq:10}) includes contributions with selected values of wave vector, proportional to the matrices \(\tilde{g}(\bm{b_m})\), and an isotropic contribution proportional to the average matrix
\(\left( \tilde{g}(b\bm{e}_{x}) + \tilde{g}(b\bm{e}_{y})\right)/2\).
In the case of rapid momentum relaxation, the isotropic contribution is
dominant as \(\tau_{i} \sim \tau \gg \tau_{p}\).
Then we may simplify the result, considering \(\tau_{\ast} \approx \tau_{p}\) and
\(\Omega\tau_{p} \ll 1\).
Upon inserting the explicit generation matrix (\ref{Eq:6}) in (\ref{Eq:10}), we get the isotropic component of the density of excitons with \(|\bm{K}| = b\) and their average spin
\begin{align}
    N &= \left( \eta^{2} + 1 \right)\frac{\tau g_{1}I_{0}}{\pi b}\delta(K - b),\nonumber\\
    M_{z} &= \frac{\tau_{1}}{\tau}\frac{2\eta}{\eta^{2} + 1}\mathcal{P}_{c}^{0},\label{Eq:11}\\
    M_{x} - iM_{y} &= \frac{1}{1 + i\Omega\tau_{2}}\frac{\tau_{2}}{\tau}\ \left( \mathcal{P}_{l}^{0} - i\frac{2\eta}{\eta^{2} + 1}\mathcal{P}_{l'}^{0} \right).\nonumber    
\end{align}
This result should be compared to the orientation of excitons at \(\bm{K}=0\), for which an equation of the kind (\ref{Eq:7}) is also applicable, except for the term corresponding to momentum relaxation \cite{Bir1973}.
In that case, given the generation matrix (\ref{Eq:4}), we obtain the exciton density \(N = \tau g_{0}I_{0}\ \delta(\bm{K})\) and average spin components
\begin{equation}
    \label{Eq:12}
    M_{z} = \frac{\tau_{1}}{\tau}\ \mathcal{P}_{c}^{0},\quad
    M_{x} - iM_{y} = \frac{1}{1 + i\Omega\tau_{2}}\frac{\tau_{2}}{\tau}\left( \mathcal{P}_{l}^{0} - i\mathcal{P}_{l'}^{0} \right).
\end{equation}
Thus, when excitons with \(|\bm{K}| = b\) are generated through the lattice of nanoparticles, the degrees of circular polarization and of linear polarization in the axes \(x',y'\) rotated by \(45^{\circ}\) with respect to the lattice are multiplied by the factor \(2\eta/(\eta^{2} + 1)\).
As in the case of \(\bm{K}=0\), circularly polarized light produces excitons whose spin is oriented along the \(z\) axis, and linearly polarized light generates excitons with spin in the perpendicular plane \(xy\).
However, we note that the lifetime \(\tau\) and spin relaxation time \(\tau_{i}\) entering equations (\ref{Eq:11}) and (\ref{Eq:12}) generally are not the same for excitons with \(|\bm{K}| = b\) or \(\bm{K}=0\).
The degrees of polarization \(\mathcal{P}_{l}^{0}\) and \(\mathcal{P}_{l'}^{0}\) enter the result (\ref{Eq:11}) differently, since we considered the generation of
excitons with wave vectors oriented along the lattice axes \(x,y\).
If the excitons with wave vectors \(\bm{K}=b(\bm{e}_{x} \pm \bm{e}_{y})\) are considered, on the contrary, the polarization degree \(\mathcal{P}_{l'}^{0}\) will appear with the factor \(1\), and the polarization degree \(\mathcal{P}_{l}^{0}\) will be multiplied by \(2\eta'/({\eta'}^{2} + 1)\), where
\(\eta' = 1 - 2b^2/q^2\).
In other words, the lattice does not alter the linear polarization in the axes that coincide with the directions of propagation of generated excitons.

\section[Luminescence]{Luminescence of hot excitons}
In previous section, we have calculated the distribution of excitons over momenta in QW and their spin density matrix resulting from continuous pumping.
Now we look into characteristics of radiation of excitons, and we apply the Fermi golden rule to calculate the probability of spontaneous photon emission \cite{Bebb1972}, since the excitonic recombination occurs via nonradiative processes predominantly.
The luminescence process, being inverse to the absorption of light, is characterized by the complex conjugate of matrix element (\ref{Eq:2}), where the external electric field \(\bm{E}_{0}\) should be replaced with the field of the ground state of selected photonic mode.
Note that photon modes in considered structure (Fig.~\ref{Fig:1}) are different from photons in homogeneous medium, but described by the wave vector \(\bm{q =}\left( \bm{q}_{\parallel},q_{z} \right)\) and polarization ($s$ or $p$) of the wave incident on the grating from the surrounding space.
Then radiation with a given frequency \(\omega\) and wave vector \(\bm{q}_{\parallel}\), detected at \(z = - \infty\), includes both modes with
\(q_{z} = \pm (q^{2} - q_{\parallel}^{2})^{1/2}\).
As shown in Fig.~\ref{Fig:3}, the grating allows the radiation of light by excitons whose
wave vector lies in the regions
\(|\bm{K} - \bm{b_m}| = |\bm{q}_{\parallel} |\bm{<}q(\omega)\),
where the radiation wave vector \(\bm{q}_{\parallel}\) can take any values inside the ``light cone''.
Up to a constant that determines the intensity of radiation, the polarization matrix of the emitted wave is given by the following expression
\begin{equation}
    \label{Eq:13}
    d_{\alpha\beta}\left( \bm{q}_{\parallel} \right) = \sum_{\bm{K},MM'} {V_{\bm{K},M;\bm{q}_{\parallel},\alpha} V_{\bm{K},M';\bm{q}_{\parallel},\beta}^{\ast}\ \rho_{MM'}(\bm{K})}.
\end{equation}
Next we consider the exciton radiation at small angle to the growth axis \(z\), such that, on the one hand, this radiation can be distinguished from the specular reflection of pump beam, and on the other hand, we can assume \(\bm{q}_{\parallel} \approx 0\) and use the equation of the kind (\ref{Eq:2}) for the matrix element of interaction operator.
Substituting into (\ref{Eq:13}) the density matrix of generated excitons with
\(|\bm{K}| = b\) determined by parameters (\ref{Eq:11}), and adding the contributions of excitons with wave vectors \(\bm{K} = \pm b\bm{e}_{x}\) and \(\pm b\bm{e}_{y}\), we
obtain the Stokes parameters of scattered radiation (which has the same frequency as the pump)
\begin{gather}
    \mathcal{P}_{l} = \frac{\tau_{2}}{\tau}\left\lbrack 1 + \left( \Omega\tau_{2} \right)^{2} \right\rbrack^{- 1}\left( \mathcal{P}_{l}^{0} - \Omega\tau_{2}\frac{2\eta}{\eta^{2} + 1}\mathcal{P}_{l'}^{0} \right),\nonumber\\
    \mathcal{P}_{l'} = \frac{\tau_{2}}{\tau}\left\lbrack 1 + \left( \Omega\tau_{2} \right)^{2} \right\rbrack^{- 1}\frac{2\eta}{\eta^{2} + 1}\left(\frac{2\eta}{\eta^{2} + 1}\mathcal{P}_{l'}^{0} + \Omega\tau_{2}\mathcal{P}_{l}^{0} \right),\nonumber\\
    \mathcal{P}_{c} = \frac{\tau_{1}}{\tau}\left( \frac{2\eta}{\eta^{2} + 1} \right)^{2}\mathcal{P}_{c}^{0}. \label{Eq:14}
\end{gather}
\begin{figure}[b]
    \centering
    \includegraphics[width=0.4\textwidth]{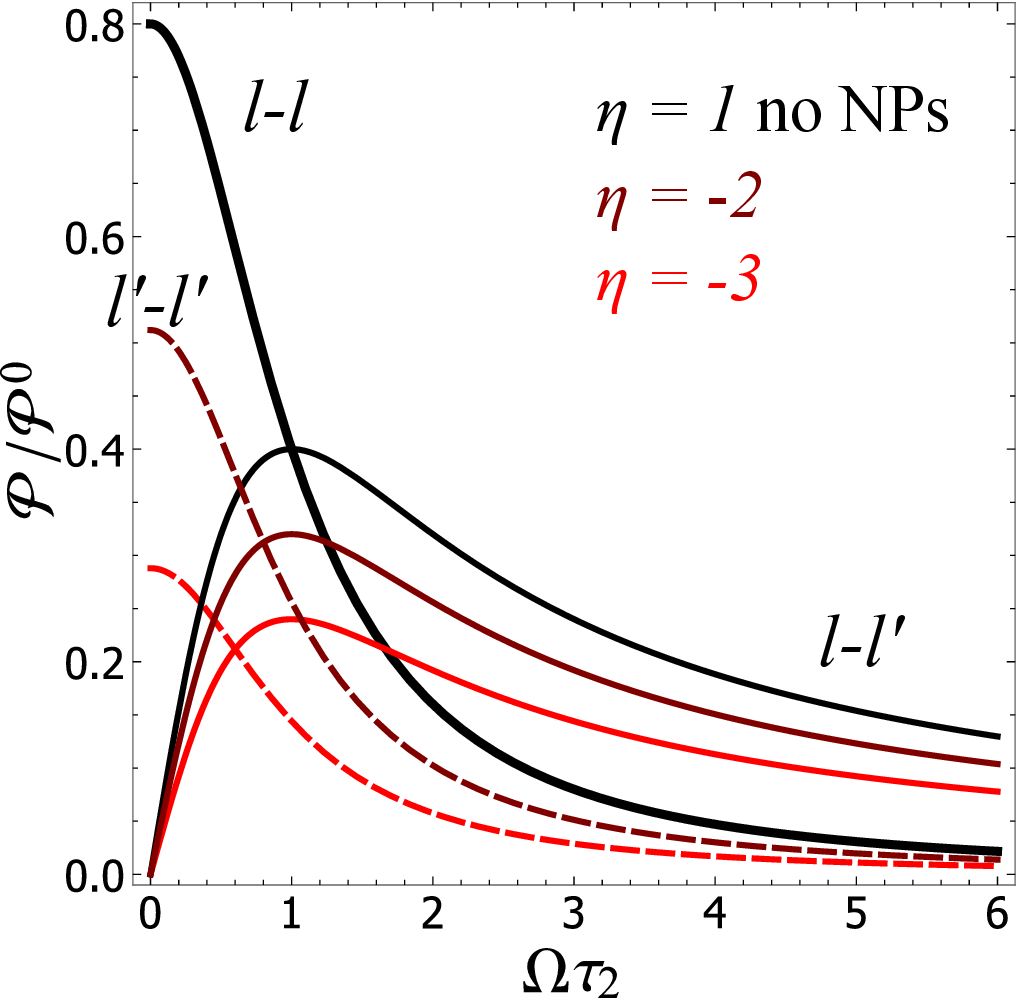}
    \caption{Ratios of the linear polarizations of luminescence and pump, measured in the \(x,y\) axes ($l$) and the \(x',y'\) axes ($l'$) rotated by \(45^{\circ}\).
    Calculated from Eq. (\ref{Eq:14}) with \(\tau_{2}/\tau\  = 0.8\).}
    \label{Fig:4}
\end{figure}
In order to describe the luminescence of excitons with \(\bm{K} = 0\), one could either substitute their spin density matrix (\ref{Eq:12}) into equation (\ref{Eq:13}), or formally set \(\eta \to 1\)
in equation (\ref{Eq:14}).
Either method yields the known result \cite{Pikus1982}:
\begin{equation}
    \label{Eq:15}
    \mathcal{P}_{l} - i\mathcal{P}_{l'} = \frac{\tau_{2}}{\tau}\frac{1}{1 + i\Omega\tau_{2}}\left( \mathcal{P}_{l}^{0} - i\mathcal{P}_{l'}^{0} \right),\quad
    \mathcal{P}_{c} = \frac{\tau_{1}}{\tau}\mathcal{P}_{c}^{0}.
\end{equation}
As seen from (\ref{Eq:14}) - (\ref{Eq:15}), the circular polarization of luminescence
reduces (compared to the pump polarization) due to the longitudinal spin relaxation, while the linear polarization is reduced by transverse spin relaxation and further suppressed by an applied magnetic field.
Additionally, in magnetic field the linear polarization of pump \(\mathcal{P}_{l}^{0}\) is partially transformed into the linear polarization of luminescence \(\mathcal{P}_{l'}\) in rotated axes (and vice versa, the polarization \(\mathcal{P}_{l'}^{0}\) is partially transformed into \(\mathcal{P}_{l}\)).

The difference in the polarized luminescence of excitons with \(\bm{K} = 0\) and \(|\bm{K}| = b\) is the lattice induced factor \(2\eta/(\eta^{2} + 1)\), which should not be too small so that the orientation of hot excitons could be observed.
For that, the lattice period \(a\) should be only several times less (2-3 times) that
the wavelength of light in semiconductor.
Fig.~\ref{Fig:4} shows the ratios of the linear polarization degrees of exciton luminescence and pump plotted as the functions of the parameter \(\Omega\tau_{2}\) (proportional to magnetic field) for several values of lattice parameter \(\eta\).
Measuring such dependences (as in Hanle effect) provides the estimations of the lifetime and spin relaxation time of excitons, but in the case \(\eta \neq 1\), when the lattice is present, these times relate to hot excitons with nonzero wave vectors.
Therefore, fabrication of a hybrid structure including a QW and a lattice of nanoparticles will make it possible to study the properties of hot excitons.

\section[Energy relaxation]{Solution of kinetic equation including the exciton energy relaxation}
\begin{figure}
    \includegraphics[width=0.35\textwidth]{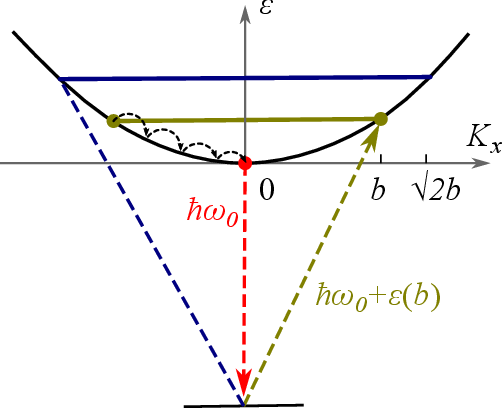}
    \caption{Schematic of excitation and radiative recombination of excitons, taking into account their energy relaxation caused by the emission of acoustic phonons.}
    \label{Fig:5}
\end{figure}

In the picture considered above, frequencies of the incident and emitted light coincide, which may hamper the observation of the polarized luminescence of hot excitons.
However, these excitons are able to lose their kinetic energy while scattering on acoustic phonons, and then emit light on a lower frequency, as shown schematically in Fig.~\ref{Fig:5}.
To describe the orientation of exciton spin in this case, it is necessary to include the energy relaxation in equation (\ref{Eq:7}) for the stationary density matrix. This can be accomplished using the equation of Fokker-Planck type \cite{Abakumov1991}, assuming that the kinetic energy of exciton changes in relatively small portions as it interacts with long-wave acoustic phonons.
Considering the spontaneous emission of phonons to be dominant over absorption, we neglect the energy diffusion term and write the equation for the exciton density matrix
\(\langle\rho\rangle\) averaged over the momentum orientations
\begin{equation}
    \label{Eq:16}
    - \frac{i\Omega}{2\hbar}\left[\sigma_{z},\left\langle \rho \right\rangle\right] - \frac{\langle\rho\rangle}{\tau} + \left( \frac{\partial\langle\rho\rangle}{\partial t} \right)_{\rm s.r.} + \frac{\partial}{\partial\varepsilon}\left( \frac{\varepsilon}{\tau_{\varepsilon}}\langle\rho\rangle\right) +\langle g \rangle = 0.
\end{equation}
Compared to equation (\ref{Eq:7}), here the almost immediate momentum relaxation
is taken into account, and the density matrix \(\langle\rho\rangle(\varepsilon)\) depends only on the kinetic energy \(\varepsilon(K) \approx \hbar^{2}K^{2}/(2\mathcal{M}_{\rm exc})\).
The new penultimate term in (\ref{Eq:16}) corresponds to the drift of excitons
towards lower kinetic energy.
We represent the exciton generation matrix averaged over momentum directions in the form \(\langle g \rangle = 1/2\; {\rm Tr}\langle g \rangle\left(1 + \bm{\sigma} \cdot \bm{M}^{0} \right)\),
and for excitation in a narrow spectral range
\({\rm Tr}\langle g \rangle = \mathcal{G}\ \delta( \varepsilon - \varepsilon_{0}) \),
so that excitons with a certain kinetic energy value \(\varepsilon_{0}\) are generated.
For example, in the case of excitons with wave vectors \(\bm{K} = \pm b\bm{e}_{x}\) and \(\pm b\bm{e}_{y}\) we have \(\varepsilon_{0} = \varepsilon(b)\), and averaging equation (\ref{Eq:6}) yields
\(\mathcal{G} = (\eta^2 + 1) g_1 I_{0}\hbar^2/(2\pi\mathcal{M}_{\rm exc})\)
and the average exciton spin at the moment of generation
\(\bm{M}^0 = \left( \mathcal{P}_{l}^{0},\ 2\eta/(\eta^2 + 1)\ \mathcal{P}_{l'}^0,\ 2\eta/(\eta^{2} + 1)\ \mathcal{P}_c^0 \right).\)

To solve equation (\ref{Eq:16}), again we use the representation of density
matrix \(\langle\rho\rangle = N/2\;(1 + \bm{\sigma} \cdot \bm{M})\),
and obtain the system of equations for concentration and mean angular momentum projections (which are now regarded as function of \(\varepsilon\))
\begin{gather}
    \label{Eq:17}
    \frac{N}{\tau} - \frac{\partial}{\partial\varepsilon} \left(\frac{\varepsilon}{\tau_\varepsilon} N \right) = \mathcal{G}\; \delta(\varepsilon - \varepsilon_{0}),\\
    \label{Eq:18}
    \frac{N M_z}{\tau_{1}} - \frac{\partial}{\partial\varepsilon }\left( \frac{\varepsilon}{\tau_\varepsilon}NM_{z} \right) = \mathcal{G} M_z^0\;\delta(\varepsilon - \varepsilon_{0}),\\
    \left\lbrack \frac{1}{\tau_{2}} + i\Omega - \frac{\partial}{\partial\varepsilon}\frac{\varepsilon}{\tau_{\varepsilon}} \right\rbrack N\left(M_x - iM_y \right) \nonumber\\
    = \mathcal{G} \left(M_x^0 - iM_y^0\right)\ \delta(\varepsilon - \varepsilon_{0}).\label{Eq:19}
\end{gather}
The last equation is valid together with its complex conjugate.
Equations with \(\delta\)-functions on the right hand side are equivalent to homogeneous ones in the interval \(0 < \varepsilon < \varepsilon_{0}\) with boundary conditions \( N(\varepsilon_{0} - 0) = \mathcal{G}\ \tau_{\varepsilon}(\varepsilon_0)/\varepsilon_0\) and \(N M_i (\varepsilon_{0} - 0) = \mathcal{G} M_i^0\ \tau_{\varepsilon}(\varepsilon_{0})/\varepsilon_{0}\),
which suggest the absence of excitons with energies \(\varepsilon > \varepsilon_{0}\).
Then the solution is
\begin{gather}
    \label{Eq:20}
    N(\varepsilon) = \mathcal{G}\;\frac{\tau_\varepsilon} {\varepsilon}\ \exp\left( - \int_\varepsilon^{\varepsilon_{0}}\frac{d\varepsilon'}{\varepsilon'}\frac{\tau_\varepsilon}{\tau} \right),\\
    \label{Eq:21}
    NM_{z} = \mathcal{G}M_{z}^{0}\;\frac{\tau_\varepsilon}{\varepsilon}\ \exp\left( - \int_{\varepsilon}^{\varepsilon_0}\frac{d\varepsilon'}{\varepsilon'}\frac{\tau_{\varepsilon}}{\tau_1} \right),
\end{gather}
and, if the notation
\(T = \int_\varepsilon^{\varepsilon_{0}} d\varepsilon'\tau_\varepsilon/\varepsilon'\)
is introduced, the transverse components of angular momentum are determined by the following equation and its complex conjugate
\begin{align}
    N\left( M_x - iM_y \right) &= \mathcal{G}\left( M_{x}^{0} - iM_{y}^{0} \right)\nonumber\\
   &\times\frac{\tau_\varepsilon}{\varepsilon} \exp\left( - \int_\varepsilon^{\varepsilon_{0}}\frac{d\varepsilon'}{\varepsilon'}\frac{\tau_{\varepsilon}}{\tau_2} - i\Omega T \right).\label{Eq:22}
\end{align}
Integrals in equations (\ref{Eq:20}) - (\ref{Eq:22}) can be estimated by assuming the lifetimes \(\tau,\tau_\varepsilon = {\rm const}(\varepsilon)\) to be independent of kinetic energy, and the spin relaxation time \(\tau_{si}(\varepsilon) \sim \varepsilon^{-1}\), which is valid in case of Dyakonov-Perel mechanism of spin relaxation \cite{Kokurin2013}.
Recollecting the notation \(\tau_i^{-1} = \tau^{-1} + \tau_{si}^{-1}\), we obtain the functions of energy
\begin{align}
    \label{Eq:23}
    N(\varepsilon) &= \mathcal{G}\;\frac{\tau_\varepsilon}{\varepsilon}\left( \frac{\varepsilon}{\varepsilon_0} \right)^{\tau_{\varepsilon}/\tau},\\
    \label{Eq:24}
    M_{z} &= M_{z}^{0}\exp\left( \frac{\tau_{\varepsilon}}{\tau_{s1}(\varepsilon)} - \frac{\tau_{\varepsilon}}{\tau_{s1}(\varepsilon_{0})} \right),\\   
    M_{x} - iM_{y} &= \left( M_{x}^{0} - {iM}_{y}^{0} \right)\nonumber\\
    &\times \exp\left( \frac{\tau_{\varepsilon}}{\tau_{s2}(\varepsilon)} - \frac{\tau_{\varepsilon}}{\tau_{s2}\left( \varepsilon_{0} \right)} \right)\exp( - i\Omega T).\label{Eq:25}
\end{align}
Therefore, as excitons lose they kinetic energy, their spin component \(M_z\) decreases exponentially, the steeper the longer the energy
relaxation time \(\tau_\varepsilon\).
Transverse spin components diminish as well when the exciton energy decreases, and additionally transform into each other due to the spin precession with frequency \(\Omega\) in applied magnetic field.
However, in contrast to result (\ref{Eq:11}) obtained in absence of energy relaxation, equation (\ref{Eq:25}) lacks the factor \((1 + i\Omega\tau_2 )^{-1}\) that ensures the decreasing dependence of transverse spin components on magnetic field.
It should be emphasized that equations (\ref{Eq:11}) and (\ref{Eq:23}) - (\ref{Eq:25}) relate to essentially different pictures.
The first case considers the polarization of excitons that have not lost their energy yet, while the second case refers to the excitons with lower kinetic energy that originate exactly from the relaxation process.

Some excitons, having lost their kinetic energy by emitting phonons, end
up near the bottom of the Brillouin zone and emit light at frequency
\(\omega\approx\omega_0\).
According to equation (\ref{Eq:13}), the polarization of luminescence of these excitons, observed along the \(z\) axis, is equal to the average spin components that are given by equations (\ref{Eq:24}) and (\ref{Eq:25}) at \(\varepsilon \rightarrow 0\).
Thus, the nanoparticle grating makes it possible to generate hot excitons with absorbed photon energies \(\hbar\omega = \mathcal{E}_{\rm exc}(\bm{b_m})\) and detect the luminescence of thermalized excitons with \(\bm{K} = 0\), which provides the information on the kinetics of relaxation processes.

\section{Conclusion}

The paper theoretically considers the optical alignment of hot excitons
in a QW, which are generated with the near field of a grating of metal
nanoparticles embedded in semiconductor heterostructure. A model is
developed to consider the effect of the grating on the exciton spin
orientation and polarization of exciton luminescence. It is demonstrated
that gratings of a period several times smaller than the wavelength of
exciting light preserve the correlation between polarizations of the
exciton luminescence and the pumping beam. This enables the experimental
determination of the lifetime and spin relaxation time of hot excitons
via the measurement of the degrees of circular and linear polarization
of secondary radiation in applied magnetic field. Apart from
quasi-resonant luminescence, due to the relaxation of exciton energy the
radiation occurs on the frequencies shifted from the excitation line
towards lower energies. The dependence of luminescence polarization of
thermalized excitons on magnetic field in Faraday geometry provides the
insight into kinetics of relaxation process.

\begin{acknowledgments}
This work was supported by Russian Science Foundation, project \textnumero~22-12-00139. 
\end{acknowledgments}


\bibliography{References}

\end{document}